\documentclass[twocolumn,floatfix,amsmath,amssymb,prb,showpacs,superscriptaddress,hyperref]{revtex4-2}
\usepackage{graphicx}
\usepackage{color}
\usepackage{bbold}
\usepackage{tikz}
\usepackage{dcolumn}
\usepackage{bm,listings}
\usepackage{dsfont}
\usepackage{ulem}
\newcommand{\rmi}{{\rm i}}
\newcommand{\rme}{{\rm e}}
\newcommand{\rmd}{{\rm d}}
\usepackage{amsfonts}
\begin{document}

\title{Landau-Zener transitions through a pair of higher order exceptional points}
\date{\today}
\author{Rishindra Melanathuru}\author{Simon Malzard}\author{Eva-Maria Graefe}
\address{Department of Mathematics, Imperial College London, London, SW7 2AZ, United Kingdom}

\begin{abstract}

Non-Hermitian quantum systems with explicit time dependence are of ever-increasing importance. There are only a handful of models that have been analytically studied in this context. Here, a PT-symmetric non-Hermitian $N$-level Landau-Zener type problem with two exceptional points of $N$-th order is introduced. The system is Hermitian for asymptotically large times, far away from the exceptional points, and has purely imaginary eigenvalues between the exceptional points. The full Landau-Zener transition probabilities are derived, and found to show a characteristic binomial behaviour. In the adiabatic limit the final populations are given by the ratios of binomial coefficients. It is demonstrated how this behaviour can be understood on the basis of adiabatic analysis, despite the breakdown of adiabaticity that is often associated with non-Hermitian systems.
\end{abstract}

\maketitle

\section{Introduction}
Quantum systems with explicitly time-dependent Hamiltonians play an important role in many experimental applications. If the parameters change sufficiently slowly in time, the adiabatic theorem states that there are no transitions of population between the different instantaneous eigenstates. When the speed of parameter change is not slow compared to characteristic scales related to the energy gaps between different instantaneous eigenstates, transitions between the states occur. Fortunately, many realistic situations can be well described by simplified models. Undoubtedly the most important of these models is the Landau-Zener (or more accurately Landau-Zener-St\"uckelberg-Majorana) model. It describes the transfer of population between two-levels driven linearly through an avoided crossing \cite{Landau1932,Zener1932,Majorana1932,Stuckelberg1932}, modelled by the time-dependent Schr\"odinger equation
\begin{equation}
\rmi\begin{pmatrix}\dot\psi_1\\ \dot\psi_2\end{pmatrix}=\begin{pmatrix}-\alpha t & v\\
v & \alpha t\end{pmatrix}\begin{pmatrix}\psi_1\\ \psi_2\end{pmatrix}.
\label{eqn:LZ_dyn}
\end{equation}
Assuming the system at time $t\to -\infty$ fulfills $|\psi_1|=1$ and $|\psi_2|=0$, i.e. $|\psi(t\to-\infty)\rangle\propto|+\rangle$, the probability $|\psi_1(t\to+\infty)|^2$, to find the system in the initial state at time $t\to+\infty$ is given by the Landau-Zener-St\"uckelberg-Majorana (LZSM) formula
\begin{equation}
\label{eqn:PLZ}
P_{LZ}=\rme^{-\frac{\pi v^2}{\alpha}}.
\end{equation}
This describes the probability of a transition between two instantaneous (or \textit{adiabatic}) states for time-dependent Hamiltonians in the vicinity of an avoided crossing of energy levels, with velocity $\alpha$ of parameter variation. In the adiabatic limit $\alpha\to 0$ this probability tends to zero, and a system follows the instantaneous states, as expected from the adiabatic theorem. 

While the LZSM probability is derived from the asymptotic behaviour of this idealised system, the result is much more robust. In practice, non-adiabatic transitions happen close to avoided energy crossings, and the behaviour away from these crossing plays no significant role. Thus the results accurately describe the transition probabilities in a large number of physical situations which in the neighbourhood of an avoided crossing of two adiabatic eigenstates can be approximated by (\ref{eqn:LZ_dyn}).

Systems with more than two relevant eigenstates with non-trivial avoided crossing scenarios are harder to treat analytically, in particular due to interference effects following multiple transitions. A number of important model systems can be solved fully and provide important insights into more general behaviour \cite{Demkov1968,Brundobler1993,Ostrovsky1997,Demkov2001,Sinitsyn_2017,Shytov2004,Yuzbashyan2018,Patra2015,Sinitsyn2004,Chen2018}. In particular, higher dimensional Hamiltonians that are elements of an $su(2)$ algebra provide model systems that show non-trivial population transfer, but can be solved using algebraic and group theoretical ideas on the basis of a corresponding $2\times 2$ model \cite{Hioe1987}. In fact, already in Majorana's original paper on the problem, higher dimensional realisations were considered \cite{Majorana1932,BlochRabi1945}. 

In recent decades there has been an ever growing interest in quantum systems described by non-Hermitian Hamiltonians, that arise naturally in the context of dissipation, scattering and losses (see, e.g. \cite{Christodoulides2018,Nimrod2011,Bender2018} and references therein). In particular non-Hermitian PT-symmetric systems that possess an anti-linear symmetry that can be interpreted as a balance of gain and loss, are leading to exciting new developments and applications \cite{El-Ganainy2018,Longhi2018}. The eigenvalues of non-Hermitian Hamiltonians are typically complex and, more importantly, their eigenvectors are in general not orthogonal to each other. This phenomenon is most pronounced at what is known as exceptional points in the parameter space, at which two or more of the eigenvectors coalesce and the Hamiltonian is not diagonalisable, but similar to a Jordan normal form (see, e.g. \cite{Heiss2012} and references therein). An exceptional point at which $N$ eigenfunctions coalesce is referred to as an exceptional point of order $N$. 

Instantaneous energies cross at exceptional points, and in their neighbourhood the eigenvalues depend on the parameters in a characteristic non-analytic fashion. In particular, $N$-th order exceptional points often appear as $N$-th root branch points in the instantaneous energies of a system. However, the eigenvalue perturbations around an $N$-th order exceptional point can follow more complicated patterns in specific cases \cite{Demange2011,Alexei2003,Ma1998} and can be understood through analysis of the Puiseux expansion \cite{Kato2013,Graefe_2008}. The non-analytic nature of the energies as functions of the parameters, combined with the influence of the non-vanishing imaginary parts causing relative exponential decays and growths, leads to unusual behaviour in the neighbourhood of exceptional points when parameters are varied in time. 
This unusual behaviour, in particular in the adiabatic regime, has been highlighted by recent theoretical \cite{Graefe2013,Berry2011,Uzdin2011,Lefebvre2009,Mailybaev2005} and experimental \cite{Yoon2018,Doppler2016,Zhang2018,Zhang2019,Liu2020} work associated with encircling exceptional points. Specifically higher order exceptional points have been explored in PT-symmetric systems in  \cite{Wimmer_2015,Lakshmanan_2021,Bergholtz_2021,Ge_2015,Peng_2020,Yoshida_2021} and in non-Hermitian systems without PT-symmetry \cite{Capolino_2017,El-Ganainy_2020,Nori_2017}.

There have been various investigations of non-Hermitian generalisations of the LZSM model \cite{Vitanov1997,Graefe2006,Reyes2012,Boyan2013,Garmon2014,Akulin1992} which involve exceptional points of order two \cite{Xu_2021,Fratalocchi_2006,Wang2022} and higher \cite{xia_2021,Lakshmanan_2021}. In particular the situation where the coupling between the two-levels in the standard LZSM scenario is made imaginary, described by the time-dependent Hamiltonian
\begin{equation}
\hat{H}^{(2\times2)} =  \begin{pmatrix} -\alpha t & i\gamma \\ i\gamma & \alpha t \end{pmatrix}
\label{eqn:H2by2}
\end{equation}
is considered in \cite{Longstaff2019,Shen2019}. 

This system is $PT$-symmetric for all times. It has imaginary eigenvalues for small times, between a pair of exceptional points at $t^{(\pm)}_{EP}=\pm \frac{\gamma}{\alpha}$. Asymptotically for larges times $t\to\pm\infty$ it reduces to the same Hermitian diagonal limit as the original LZSM Hamiltonian. The presence of the imaginary non-reciprocal coupling changes the characteristics of the transitions, and in the adiabatic limit, $\alpha\to 0$, the final state equally populates the two levels, irrespective of the initial state. 

In the present paper we generalise these results to a non-Hermitian $N$-level system driven through a pair of exceptional points. In particular, we consider the model
\begin{equation}
\hat{H} = -2\alpha t\hat{J}_{z} + 2i\gamma \hat{J}_{x},
\label{eqn:genNHt}
\end{equation}
where $\alpha$ and $\gamma$ are real and positive constants, and $\hat{J}_{x}, \hat{J}_{y}, \hat{J}_{z}$ are the standard quantum angular momentum operators, i.e., generators of the $su(2)$ algebra. The model has two $N$-th order exceptional points at which the matrix is similar to a full Jordan block. For large $t$ the model again approximates the Hermitian limit $\hat H\sim-2\alpha t\hat{J}_{z}$, and transition probabilities are well defined. Using the $SL(2)$ structure of the model we derive the full set of transition probabilities between the asymptotic eigenstates of the $N$-dimensional model from the $2\times2$ realisation. The algebraic structure and the PT-symmetry lead to a square root unfolding of the eigenvalues around the $N$-th order exceptional point \cite{Graefe_2008}. In the adiabatic limit this leads to a non-trivial re-distribution of populations following a binomial pattern. Interestingly, this pattern can be understood on the grounds of an adiabatic argument, arising from the geometry of the eigenvector system close to the exceptional point. 

The paper is organised as follows. In section \ref{sec:HermitianSU2} we review the standard LZSM model and summarise how the $N$ level generalisation can be solved using its group structure \cite{Hioe1987}. We then move on to the discussion of the non-Hermitian system (\ref{eqn:genNHt}) in section \ref{sec_nHerm}, where we derive the full set of velocity-dependent transition probabilities between the different asymptotic states. We provide a derivation for the non-trivial adiabatic limiting behaviour, based on the adiabatic eigenvector structure in section \ref{sec:AdiabaticTheorem}. We close with a short summary in section \ref{sec:sum}. An appendix provides some details on how the $SL(2)$ structure is used to deduce an $N$-dimensional representation of a group element from the $2\times2$ representation. For convenience we use dimensionless units throughout the paper with $\hbar=1$.

\section{Landau-Zener-St\"uckelberg-Majorana Transitions in N-level, $SU(2)$, Hermitian Hamiltonians}
\label{sec:HermitianSU2}

The original LZSM model, described by the time-dependent Schr\"odinger equation (\ref{eqn:LZ_dyn}) considers two quantum states with an energy difference that changes linearly in time, coupled by the coupling constant $v\in\mathds{R}$. For vanishing coupling $v=0$, the Hamiltonian of equation (\ref{eqn:LZ_dyn}) is diagonal with the instantaneous eigenvalues $\lambda_{1,2}=\pm\alpha t$, and eigenstates $|+\rangle, |-\rangle$ of the $\hat{\sigma}_{z}$ operator. At $t=0$ it has a diabolical point \cite{Berry2011} where the two eigenvalues degenerate. 
For non-zero coupling between the two levels, $v\neq 0$, the energies as a function of the time display an avoided crossing at $t=0$ with energy gap $\Delta = 2|v|$. For $v\neq0$ the instantaneous energies are given by 
\begin{equation}
\lambda_{\pm}=\pm\sqrt{\alpha^2t^2+v^2},
\end{equation}
with corresponding eigenstates 
\begin{equation}
\phi_{\pm} \propto\begin{pmatrix} \lambda_{\pm}  - \alpha t  \\ v \end{pmatrix}.
\end{equation}
In the asymptotic limit that $t\rightarrow \pm\infty$ these are the same eigenstates as those of the uncoupled system. The LZMS probability (\ref{eqn:PLZ}) of finding the system in the initial eigenstate after parameter variation from $t\to-\infty$ to $t\to+\infty$, corresponds to the probability of transition between the instantaneous eigenstates that interchange at the avoided crossing, hence it is often referred to as a ``transition probability''. One can further ask about the probability of the system being found in the state $|+\rangle$ at $t\to+\infty$ when starting from the state $|-\rangle$ at $t\to-\infty$, and of course the equivalent questions when starting from the state $|+\rangle$. Due to the unitarity of the problem, these four probabilities are all determined by the LZSM probability (\ref{eqn:PLZ}). They can be summarised in a matrix $M$ of transition probabilities between the asymptotic eigenstates as 
\begin{equation}
M= \begin{pmatrix} e^{-\frac{\pi v^2}{\alpha}} & 1-e^{-\frac{\pi v^2}{\alpha}}  \\  1-e^{-\frac{\pi v^2}{\alpha}} & e^{-\frac{\pi v^2}{\alpha}}\end{pmatrix}.  
\label{eqn:P2}
\end{equation}
The transition probabilities starting from $|+\rangle$ in dependence on $\alpha$ are depicted in the right panel of figure \ref{fig:LZ2x2} for $v=1$.

Of course, in many realistic situations more than two levels are involved in avoided crossings. Viewing the Hamiltonian in the LZSM problem as a two-dimensional representation of an $su(2)$ Lie algebra element, it is natural to consider the $N$-dimensional model 
\begin{equation}
\hat{H}(t) = -2\alpha t \hat{J}_{z}+2v\hat{J}_{x}, 
\label{eqn:HhermN}
\end{equation}
where $\hat J_z$ and $\hat J_x$ are $N$-dimensional representations of angular momentum operators, and 
$v$ and $\alpha$ are the same real and positive constants as in the $2\times 2$ case. Perhaps not surprisingly, the transition probabilities for the $N$-dimensional problem (\ref{eqn:HhermN}) are fully determined by the solutions of the $2\times 2$ case, due to the $su(2)$ algebraic structure \cite{Hioe1987}. Let us now proceed to review the solution of the $N$ dimensional problem using this idea. 

\begin{figure}
\includegraphics[width=0.5\textwidth]{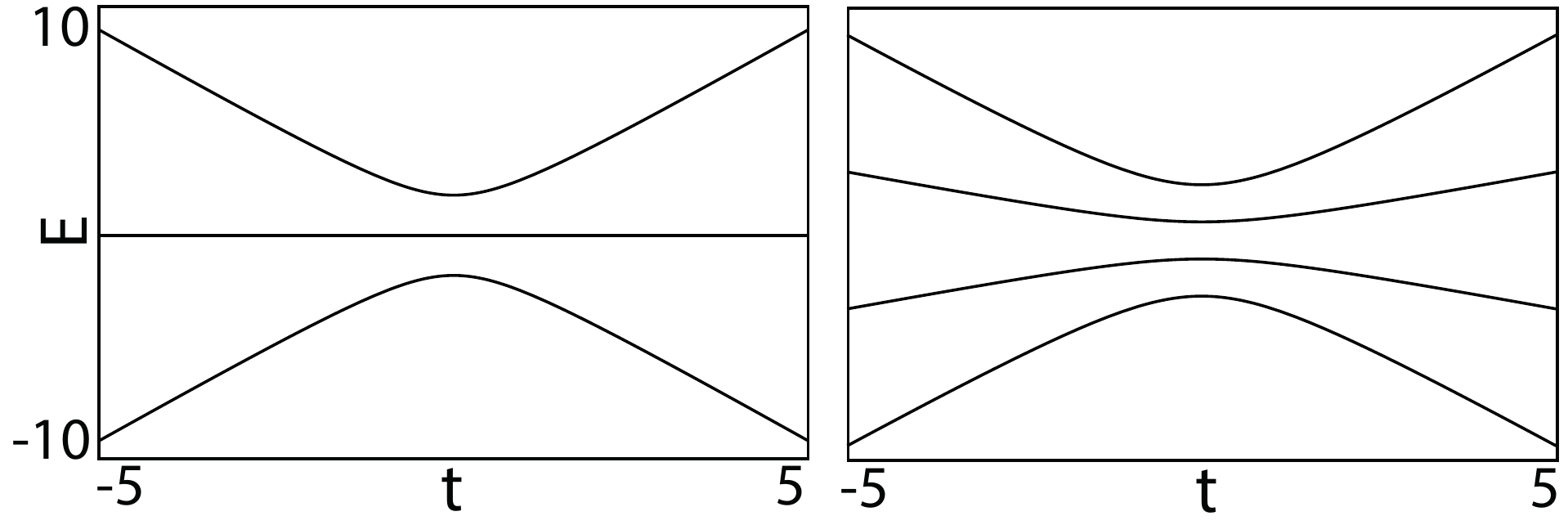}
\caption{Eigenvalues as a function of $t$ of the $3$ level (left panel) and $4$ level (right panel) Hamiltonians given in equation (\ref{eqn:HhermN}) for $N=3$ and $N=4$ respectively, where $\alpha=v=1$.}
\label{fig:3x3_4x4_Hermitian}
\end{figure}

The $N$-dimensional Hamiltonian (\ref{eqn:HhermN}) can be diagonalised by the simple rotation
\begin{equation}
\hat{H}= 2\lambda\,\hat\Phi_{herm}\hat{J}_{z}\hat \Phi_{herm}^{-1},   
\label{eqn-H_herm_diag}
\end{equation}
with 
\begin{equation}
\label{eqn:Phi_herm}
\hat\Phi_{herm}=e^{\rmi\, \phi\hat J_y},
\end{equation}
where 
\begin{equation}
\cos(\phi)=-\frac{\alpha t}{\lambda},\quad \sin(\phi)=-\frac{v}{\lambda},
\end{equation}
and 
\begin{equation}
\lambda=\sqrt{\alpha^2 t^2+v^2}.
\end{equation} 
Thus, the eigenvalues are given by multiples of those of the $2\times 2$ system, $2m\lambda$ with $m=J,\,J-1,\dots,-J+1,\,-J$, where $J=\frac{N-1}{2}$ is the angular momentum quantum number. Figure \ref{fig:3x3_4x4_Hermitian} depicts the eigenvalues as a function of time for the cases $N=3$ and $N=4$, for $\alpha=1=v$, displaying an avoided crossing of all $N$ levels around $t=0$. Note that while the avoided crossing scenario looks similar, this model is different from what is known as the \textit{bow tie} model \cite{Carroll_1986,Hioe_1986,Ostrovsky1997,Demkov2001}.

The corresponding eigenvectors are given by
\begin{equation}
|\phi^{herm}_j\rangle=\hat \Phi_{herm}|j\rangle,
\label{eqn:phi_m}
\end{equation}
where $|j\rangle$ denotes the eigenvectors of $\hat J_z$ corresponding to the eigenvalues $J-j$. That is, $j$ goes from $0$ to $N-1$. For $t\to\pm\infty$  we have $\hat H\to \mp\alpha t\hat J_z$, and the asymptotic eigenvectors are given by those of $\hat J_z$, as expected. 

The LZSM transition probabilities between the asymptotic eigenstates $M_{jk}$ are given by the squared moduli of the matrix elements of the time evolution operator $\hat U(t,t_0)$ from $t_0=-\infty$ to $t=+\infty$. Since the Hamiltonian is an element of the $su(2)$ Lie algebra, the time evolution operator $\hat U$ is an element of the group $SU(2)$. Thus, the $N$-dimensional representation can be deduced from the $2$-dimensional one. 

In fact, the transition probabilities only depend on the squared moduli of the elements of the $2\times2$ time evolution operator, that is, the transition probabilities in the $2\times 2$ case. This can be seen as follows. In general, the matrix elements of a group element $D$ of the group $SL(2)$ (the complexification of the group $SU(2)$) in $N$ dimensions are given in terms of the $2\times2$ representation, which for simplicity we shall denote by the lower case letter $\hat d$, as
\begin{align}
\nonumber
D^{(N)}_{jk}\!=&\!\begin{pmatrix}n\!\\j\!\end{pmatrix}^{\!\!\frac{1}{2}}\!\!\!\begin{pmatrix}n\!\\k\!\end{pmatrix}^{\!\!\!-\frac{1}{2}}\!\!\!\sum_{l=l_{\rm min}}^{l_{\rm max}}\!\!\!\begin{pmatrix}n\!-\!j\\k\!-\!l\end{pmatrix}\!\!\begin{pmatrix}j\\l\end{pmatrix}\!d_{11}^{n\!-\!j\!-\!k\!+\!l}d_{12}^{k\!-\!l}d_{21}^{j\!-\!l}d_{22}^{l},
\\
=&\!\!\!\sum_{l=l_{\rm min}}^{l_{\rm max}}\!\!\!\!\tfrac{\sqrt{k!(n\!-\!k)!(n\!-\!j)!j!}}{(n\!-\!j\!-\!k\!+\!l)!(k\!-\!l)!(j\!-\!l)!l!}d_{11}^{n\!-\!j\!-\!k\!+\!l}d_{12}^{k\!-\!l}d_{21}^{j\!-\!l}\!d_{22}^{l},
\label{eqn:D}
\end{align}
where
\begin{eqnarray}
l_{\rm min}&=&\text{max}(k-(n-j),0),\\
l_{\rm max}&=&\text{min}(k,j),
\end{eqnarray}
 and the $d_{jk}$ denote the matrix elements of $\hat d$, where for ease of notation the indices run from zero to $n=N-1$. This can be deduced for example from the action of the group element on the basis of coherent states of the complex projective Hilbert space $\mathds{CP}^n$, as summarised in appendix \ref{app:SL2}. 

In the $2\times 2$ case, due to its unitarity, the time-evolution operator is of the form
\begin{equation}
\hat{U}(t=\infty,t=-\infty) = \begin{pmatrix} a& b \\ -b^{*} & a^{*} \end{pmatrix},
\label{eqn:Utt}
\end{equation}
with $|a|^2+|b|^2=1$. The matrix elements of $\hat U$ in the $N$-dimensional representation can be deduced from the $2\times 2$ representation using equation (\ref{eqn:D}).
%
Here, we are interested in the transition probabilities between the diabatic states (i.e. the eigenstates at $t\to\pm\infty$), rather than the full time-evolution operator. The transition probability between states $k$ and the state $j$ is given by $M_{jk}=|\langle j|\hat U|k\rangle|^2=|\hat U_{jk}|^2$. Due to the unitarity of $\hat U$ these are functions of only the squared moduli of the matrix elements of the $2\times 2$ representation, $A=|a|^2$ and $B=|b|^2$, i.e. the transition probabilities of the two-level case. 

Using equation (\ref{eqn:D}) to deduce $U^{(N)}_{jk}$, after some algebra we find the transition matrix elements
\begin{equation}
M_{jk}=\!\!\sum_{l,l'}(-1)^{l+l'}c_{ll'} A^{n-j-k+l+l'}B^{k+j-(l+l')},
\label{eqn:M}
\end{equation}
with
\begin{equation}
\label{eqn:coeffsM}
c_{ll'}=\begin{pmatrix}k\\l \end{pmatrix}\begin{pmatrix}n-k\\j-l \end{pmatrix}\begin{pmatrix}n-j\\k-l' \end{pmatrix}\begin{pmatrix}j\\l' \end{pmatrix}.
\end{equation}

\begin{figure}
\includegraphics[width=0.5\textwidth]{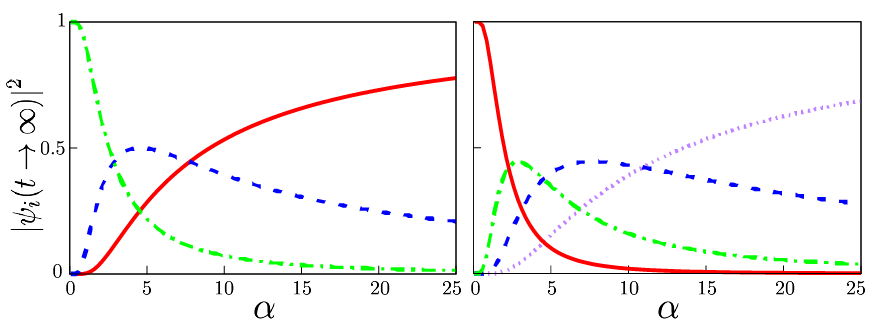}
\caption{Transition probabilities as a function of $\alpha$ for $v=1$ for the $3$ level (left) and $4$ level (right) Hermitian Hamiltonians given by equation (\ref{eqn:M_0m}). The green dot-dashed, blue dashed, red solid (and purple dotted) lines correspond to $j=0,1,2(,3)$ respectively for the $3$ level ($4$ level) Hamiltonian.}
\label{fig:Herm_alpha_t}
\end{figure}

In particular the transition probabilities between the first diabatic state and any other states, $M_{j0}$, only have one contributing term that is $l=l^{'}=0$ and thus $M_{j0}$ is simply given by the binomial expression
\begin{equation}
M_{j0}=\begin{pmatrix}n\\j\end{pmatrix}A^{n-j}B^j,
\label{eqn:M_0m}
\end{equation}
with $A=\rme^{-\frac{\pi v^2}{\alpha}}$ and $B = 1-\rme^{-\frac{\pi v^2}{\alpha}}$. These transition probabilities are plotted in figure \ref{fig:Herm_alpha_t} for dimensions $3$ and $4$ as a function of $\alpha$ for $v=1$.

For intermediate velocities of parameter variation the binomial character of the transitions between different levels becomes apparent in the multi level systems as compared to the two level system. In the adiabatic limit we have $A\to 0$ and $B\to 1$ and thus $M_{0k}\to\delta_{nk}$, as expected from the structure of the instantaneous eigenvalues. In the quantum quench limit, on the other hand, we have $A\to 1$ and $B\to 0$, and thus $M_{0k}\to\delta_{0k}$. In neither of these limits can the binomial structure of the transition probabilities be seen. 

\section{The non-Hermitian model}
\label{sec_nHerm}
Let us now turn to the main topic of this paper, the non-Hermitian Hamiltonian (\ref{eqn:genNHt}), the $N$-level generalisation of the $2\times 2$ system given in equation \ref{eqn:H2by2} which we first review. 

Here $\alpha t$ and $-\alpha t$ are the on-site energies of the two states under consideration, which are linearly changed in time as in the standard LZSM model, with velocity $\alpha$, assumed to be positive here without loss of generality. The real-valued time parameter $t$ runs from $-\infty$ to $\infty$. The system is $PT$-symmetric, i.e. it fulfils $[\hat P \hat T,\hat H]=0$, where $\hat T$ is a time-reversal operator, which in the present case is given by the complex conjugation operator, i.e. $\hat T:\rmi\mapsto-\rmi$, and $\hat P=\hat\sigma_z$ is a parity operator that maps $\hat\sigma_x$ to $-\hat\sigma_x$. This $2\times2$ model has been analysed in detail in references \cite{Longstaff2019,Shen2019}. Note that the time-reversal operator $\hat T$ is an anti-linear operator and thus does not flip the sign of the parametric time $t$. 

In contrast to the Hermitian LZSM model, the system (\ref{eqn:H2by2}) has a non-reciprocal imaginary coupling between the two states, described by the real parameter $\gamma$, with $\gamma>0$. As discussed in \cite{Longstaff2019} it naturally arises as a linear approximation of the Bloch Hamiltonian in a double-periodic lattice with absorption/losses in every other lattice site, under the influence of a static force. A more direct implementation in two wave guides would be possible if non-reciprocal imaginary couplings could be achieved. This is a topic of some research effort, and might be achievable in the near future \cite{Kenta2022,Szameit2014}. 

The eigenvalues of the $2\times 2$ model (\ref{eqn:H2by2}) are given by 
\begin{equation}
\lambda_{\pm} = \pm\sqrt{\alpha^{2}t^2 - \gamma^2},
\label{eqn:2x2eigs} 
\end{equation}
and are real for times $|t|\geq\frac{\gamma}{\alpha}$. At times $t=\pm\frac{\gamma}{\alpha}$ (which we denote as $t^{\pm}_{EP}$) the system has exceptional points. For times $|t|\leq\frac{\gamma}{\alpha}$ they are purely imaginary. The two exceptional points arise from the well-known splitting of the diabolical point at $t=0$ in the Hermitian model with $\gamma=0$ \cite{Keck2003}.

The (non-orthogonal) eigenvectors can be parameterised in a similar way to the Hermitian case as
\begin{equation}
\phi_{\pm} \propto\begin{pmatrix} \lambda_{\pm} - \alpha t \\ \rmi \gamma \end{pmatrix}.
\end{equation}
In the asymptotic limit $t\to\pm\infty$ these approach the eigenvectors of $\hat\sigma_z$ as for the Hermitian model. For finite times, however, the eigenvectors are non-orthogonal, but linearly independent outside the exceptional points. At the exceptional points the Hamiltonian is not diagonalisable as the two eigenvectors coalesce to the single eigenvector 
\begin{equation}
\phi_{EP^-}\propto\begin{pmatrix}  1 \\ \rmi  \end{pmatrix}
\end{equation}
at the exceptional point at $t_{EP}^{-}=-\frac{\gamma}{\alpha}$, and 
\begin{equation}
\phi_{EP^+}\propto\begin{pmatrix}  -1 \\ \rmi  \end{pmatrix}
\end{equation}
at the exceptional point at $t_{EP}^{+}=+\frac{\gamma}{\alpha}$. 

The transition probabilities between the asymptotic eigenstates in the time-dependent system (\ref{eqn:H2by2}) can be derived following much the same route as Zener's in the solution of the original Hermitian problem, or a more general Laplace transform technique as has been employed for a number of generalised LZSM problems by Demkov et al \cite{Demkov2001}. The non-normalised transition probabilities 
\begin{equation}
M_{jk}=\left.|\psi_j(\infty)|^2\right|_{\psi_{k}(-\infty)=1}
\end{equation}
are given by \cite{Longstaff2019}
\begin{equation}
M = 
\begin{pmatrix} e^{\frac{\pi\gamma^2}{\alpha}} & e^{\frac{\pi\gamma^2}{\alpha}}-1\\ e^{\frac{\pi\gamma^2}{\alpha}}-1 & e^{\frac{\pi\gamma^2}{\alpha}} \end{pmatrix}.
\label{eqn:M2}
\end{equation}
In the non-Hermitian system the norm is not conserved in time. If we are asking for the probabilities of transition between the asymptotic (orthogonal and stable) eigenstates, these are given by the renormalised populations at $t\to\infty$ with the respective boundary conditions, i.e. 
\begin{equation}
\begin{split}
P_{jk}&=\left.\frac{|\psi_k(\infty)|^2}{\sum_l |\psi_l(\infty)|^2  }\right|_{\psi_{j}(-\infty)=1}\\
&=\frac{M_{jk}}{\sum_l M_{jl}}.
\end{split}
\end{equation}
Explicitly these transition probabilities are given by 
\begin{equation}
P_{00}=P_{11}=\frac{1}{2-e^{-\frac{\pi\gamma^2}{\alpha}}},
\label{eqn:LZ2nHerm1}
\end{equation}
and 
\begin{equation}
P_{01}=P_{10}=\frac{1-e^{-\frac{\pi\gamma^2}{\alpha}}}{2-e^{-\frac{\pi\gamma^2}{\alpha}}},
\label{eqn:LZ2nHerm2}
\end{equation}
in the $2\times 2$ case.

\begin{figure}
\includegraphics[width=0.5\textwidth]{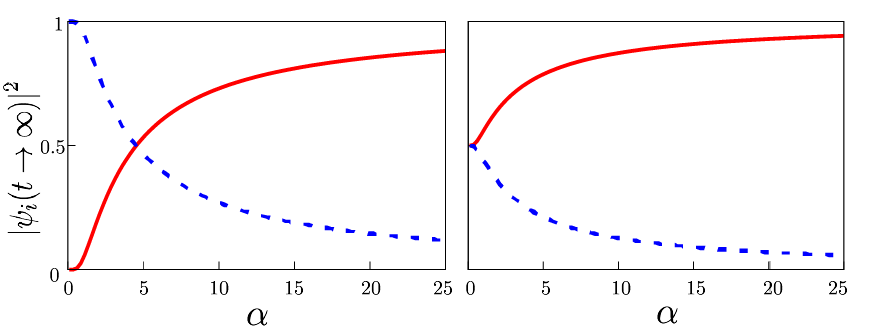}
\caption{Transition probabilities as a function of $\alpha$ for $v=1$ for the Hermitian (left) in comparison to the non-Hermitian (right) $2$ level systems. In the left panel the red solid and blue dashed lines correspond to $j=0,1$ in equation (\ref{eqn:M_0m}) and in the right panel the red solid and blue dashed lines correspond to $j=0,1$ in equation (\ref{eqn:P_0knHerm}).}
\label{fig:LZ2x2}
\end{figure}

The final population probabilities of the instantaneous basis states when the first eigenstate is initially populated in dependence on the driving rate according to equations (\ref{eqn:LZ2nHerm1}) and (\ref{eqn:LZ2nHerm2}) are depicted in the right panel of figure \ref{fig:LZ2x2}. In the quantum-quench limit of rapid parameter variation, $\alpha \gg \gamma^2$, as in the Hermitian case, the state cannot adapt in time, and remains in the original eigenstate, i.e. the probability matrix $M$ becomes diagonal. This may be intuitively understood form the fact that the two exceptional points occurring in the non-Hermitian system can be viewed as the remnants of a diabolic point which has bifurcated. It seems intuitive that in the limit of fast parameter variation the details of the crossing close to $t=0$ are not resolved by the dynamics.  

For intermediate driving rates, we observe a typical LZSM-like behaviour interpolating the adiabatic and the quantum quench limit. However, the adiabatic limit of the non-Hermitian case is crucially different from that of the Hermitian one. In the adiabatic limit, $\alpha \rightarrow 0$, we have $e^{-\frac{\pi\gamma^2}{\alpha}}\rightarrow 0$, and all $P_{jk}\to \frac{1}{2}$ \cite{Longstaff2019,Shen2019}, in contrast to the adiabatic limit of the Hermitian system, where the final state is the asymptotic state that is adiabatically connected to the initial state, and there is no population in the other eigenstate.

The non-Hermitian adiabatic behaviour is not entirely surprising, as there is no smooth continuation of the adiabatic eigenstates through the exceptional point. At the exceptional point both eigenstates are degenerate and the adiabatic eigenstates on one side cannot in any meaningful way be connected to the ones on the other side. Hence, one can argue that transitioning through the exceptional point in an adiabatic fashion the only way for the population to behave after the exceptional point is to be equal in both branches of the eigenstates.  In the model (\ref{eqn:H2by2}) the eigenvalues are complex between the two exceptional points, leading to an exponential decay of the population into the eigenstate with the positive imaginary part. In the case of an adiabatic time evolution, this means that for $|t|<|t_{EP}^{\pm}|$ the population is entirely in the eigenstate with the positive imaginary part of the energy. Once we pass the second exceptional point at $t=t^{+}_{EP}=\frac{\gamma}{\alpha}$, the indistinguishable nature of the eigenstates leads to an equal share of the final state populations of $\psi_{1}$ and $\psi_{2}$. In fact, this argument holds in the adiabatic limit, no matter what the initial conditions at $t\to-\infty$ are, and hence the final state populations are independent of initial state populations in the adiabatic limit \cite{Longstaff2019,Shen2019}. Following this argument, one may expect that the final state populations in the adiabatic limit of an $N$ level system would be equally distributed, that is $|\psi_{j}|^2=\frac{1}{N}$. As we shall uncover in what follows, however, the situation is more complicated. 

\begin{figure}
\includegraphics[width=0.5\textwidth]{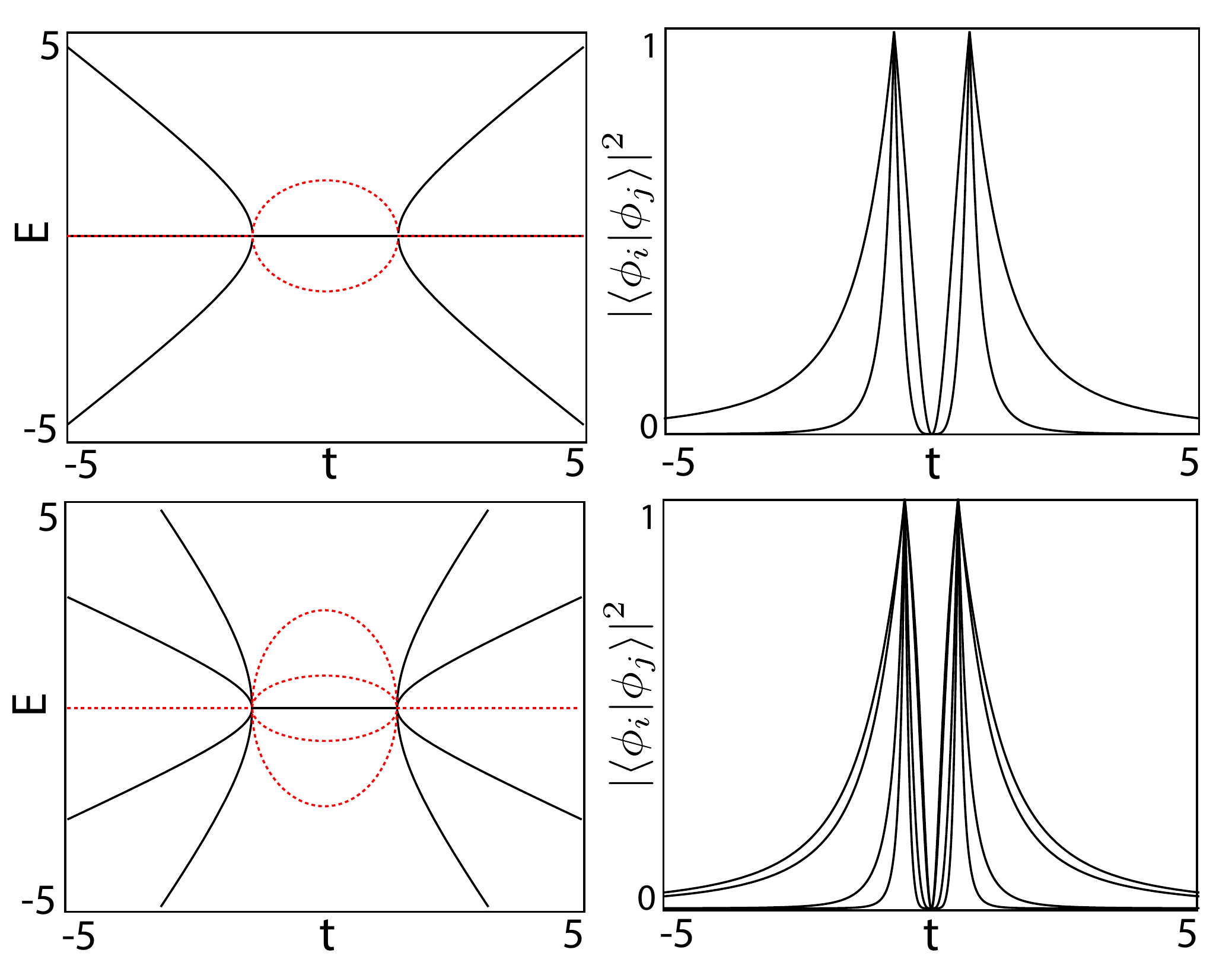}
\caption{Eigenvalues (left column)  and overlap of the eigenstates (right column) of the $3$ level (upper row) and $4$ level (bottom row) systems of equation (\ref{eqn:genNHt}) as a function of $t$, where $\alpha=\gamma=1$. Here the eigenvectors are normalised in the conventional way as $\langle \phi_j|\phi_j\rangle=1$. In the left column, the real parts of the eigenvalues are depicted by black solid lines and the imaginary parts by red dashed lines.}
\label{fig:3x3_4x4}
\end{figure}

The $N$-level generalisation we consider in the present paper, described by the Hamiltonian \ref{eqn:genNHt}, has the same overall structure as the $2\times 2$ case, but the diabolical point in the Hermitian system is of higher order and splits up into two exceptional points of order $N$. 
Away from the exceptional points the Hamiltonian is diagonalisable and can be expressed as a similarity transformation of $\hat J_z$ in analogy to the Hermitian case. It is more convenient for the calculations in the remainder of the paper, however, to express $\hat H$ as a similarity transformation of $\hat J_y$ as
\begin{equation}
\hat{H}= 2\lambda\,\hat\Phi\hat{J}_{y}\hat \Phi^{-1},   
\label{eqn-H_diagy}
\end{equation}
with 
\begin{equation}
\label{eqn:Phi_y}
\hat\Phi=e^{\rmi\frac{\pi}{2}\frac{1}{\lambda}(\alpha t\hat J_x+\rmi\gamma\hat J_z)},
\end{equation}
and 
\begin{equation}
\lambda=\sqrt{(\alpha t)^2-\gamma^2}.
\end{equation} 
As in the Hermitian case, the eigenvalues of $\hat H$ are simply given by multiples of those of the two-level system, as $E_j=(J-j)\lambda$ with $j=0,1,\dots,\,n$, and where $J=\frac{n}{2}$ is the angular momentum quantum number. The eigenvalues of the $3\times 3$ and $4\times 4$ systems are depicted in dependence on $t$ for $\alpha=1=\gamma$ in the left panel of figure \ref{fig:3x3_4x4}. Just as in the $2\times2$ system the asymptotic behaviour for large absolute values of $t$ is dominated by the $\hat J_z$ term and is thus identical to the behaviour of the Hermitian model (\ref{eqn:HhermN}).
The corresponding eigenvectors can be expressed as 
\begin{equation}
|\phi_j\rangle\propto\hat \Phi|j_y\rangle
\label{eqn:phi_j}
\end{equation}
where $|j_y\rangle$ are the eigenstates of $\hat J_y$, corresponding to the eigenvalues $J-j$.  

It is well-known that the eigenvectors of non-Hermitian operators are in general not orthogonal, $\langle\phi_j|\phi_k\rangle\neq \delta_{jk}$. In the present case, all eigenvectors coalesce into a single eigenvector at the exceptional points at $t_{EP}^{\pm}=\pm\frac{\gamma}{\alpha}$, while the system becomes approximately Hermitian in the asymptotic limits $t\to\pm\infty$, where the eigenvectors become orthogonal. In the right panels in figure \ref{fig:3x3_4x4} the overlap between pairs of eigenvectors belonging to different eigenvalues is plotted as a function of time for the $3\times3$ and $4\times 4$ case, when the eigenvectors are normalised in the conventional way. The exceptional points of higher order are clearly apparent at the sharp peaks at $t_{EP}^{\pm}=\pm\frac{\gamma}{\alpha}$, where the eigenstates become parallel and thus have maximal overlap. Note that the different eigenvectors unfold at different rates away from the exceptional points towards the asymptotic limit where they become orthogonal.

At the exceptional points the Hamiltonian cannot be diagonalised, which manifests in a divergence of the exponent in (\ref{eqn:Phi_y}), with $\lambda=0$. Instead, $\hat H(t_{EP}^{\pm})$ is similar to a Jordan block . Explicitly, $\hat H(t_{EP}^{+})$ can be transformed into $\hat J_{+}$ by a unitary transformation
\begin{equation}
\hat H(t_{EP}^+)=:\hat H_{EP}=\rmi\gamma(\hat J_x+\rmi \hat J_z)= \rmi\gamma e^{-\rmi\frac{\pi}{2} \hat{J}_{x}}\hat{J}_{+} e^{i\frac{\pi}{2}\hat{J}_{x}}.
\end{equation}
Its single eigenstate is given by 
\begin{equation}
|\phi_{EP}\rangle=e^{-i\frac{\pi}{2}\hat{J}_{x}}|0\rangle,
\end{equation}
where $|0\rangle$ denotes the eigenstate of $\hat J_z$ with the largest eigenvalue, $J$, which is also the single eigenstate of $\hat J_+$. The rotation $e^{-i\frac{\pi}{2}\hat{J}_{x}}$ turns this into the lowest eigenstate of $\hat J_y$,
\begin{equation}
|\phi_{EP}\rangle=|n_y\rangle,
\end{equation}
belonging to the eigenvalue $-J$.

We now turn to calculate the transition probabilities in the $N\times N$ case from the $2\times 2$ system. While the time-evolution operator is not unitary in this case, and is not of the form (\ref{eqn:Utt}) for the $2\times2$ case, the $PT$-symmetry of the system leads to different constraints, that again allows us to deduce the square moduli of the time evolution operator elements for the $N\times N$ case from the $2\times 2$ one without knowledge of the phases, as in the Hermitian case. Specifically, our Hamiltonian is $PT$-symmetric with $\hat P=\hat \sigma_z$, i.e. $\hat H=\hat P\hat H^*\hat P$, and is a trace-free element of the complexified $su(2)$ algebra, $sl(2)$. As a consequence, the time-evolution operator is an $SL(2)$ element which fulfils the symmetry condition $\hat U^{-1}=\hat P\hat U^*\hat P$, and, in the $2\times 2$ case can be parameterised as  
\begin{equation}
\hat U= \begin{pmatrix} a& b \\ b^{*} & a \end{pmatrix},
\label{eqn:Utt_PT}
\end{equation}
with $a\in\mathds{R}$, and $a^2-|b|^2=1$. Comparison with (\ref{eqn:M2}) shows that for the time evolution from $t\to-\infty$ to $t\to+\infty$ we have 
\begin{equation}
a^2= e^{\frac{\pi\gamma^2}{\alpha}},
\end{equation}
and 
\begin{equation}
|b|^2=e^{\frac{\pi\gamma^2}{\alpha}}-1.
\end{equation}
For the $N\times N$ case, applying the general formula (\ref{eqn:D}) yields
\begin{equation}
M_{jk}=\!\!\sum_{l,l'}c_{ll'} A^{n-j-k+l+l'}B^{k+j-(l+l')},
\label{eqn:M_2}
\end{equation}
with $A=a^2$ and $B=|b|^2$, and $c_{ll'}$ given by the product of binomial coefficients in equation (\ref{eqn:coeffsM}). 
The actual transition probabilities between the asymptotic eigenstates are then given by 
\begin{equation}
P_{jk}=\frac{M_{jk}}{\sum_l M_{lk}}.
\end{equation}
In particular, as in the Hermitian case, the transition probabilities from the lowest asymptotic eigenstate at $t\to-\infty$ is given by 
\begin{equation}
P_{j0}=\frac{M_{j0}}{\sum_l M_{l0}},
\end{equation}
with $M_{j0}=\begin{pmatrix}n\\j\end{pmatrix}A^{n-j}B^j$, that is 
\begin{equation}
P_{j0}=\begin{pmatrix}n\\j \end{pmatrix}\frac{A^{n-j}B^j}{(A+B)^n},
\label{eqn:P_0knHerm}
\end{equation}
where we observe a similar binomial structure as in the Hermitian system (\ref{eqn:M_0m}). 

\begin{figure}
\includegraphics[width=0.5\textwidth]{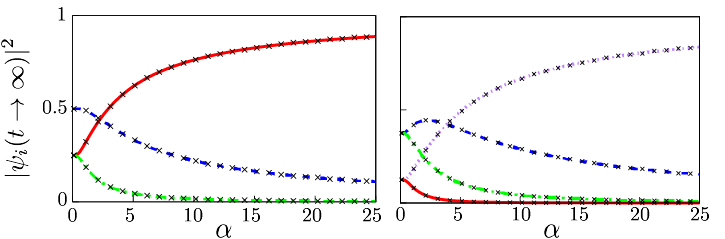}
\caption{Final state populations for the $3$ level (lower left panel) and $4$ level (lower right panel) Hamiltonians in dependence on $\alpha$ for $\gamma=1$, where the red solid, blue dashed and green dot-dashed lines (and purple dotted line) correspond to the transition probabilities given in equation (\ref{eqn:P_0knHerm}) for $j=0,1,2(,3)$ respectively for the $3$ level ($4$ level) Hamiltonian and the black crosses correspond to numerical simulations.}
\label{fig:alphat}
\end{figure}

The transition probabilities (\ref{eqn:P_0knHerm}) are depicted as a function of $\alpha$ for $\gamma=1$ for the $3\times3$ and the $4\times4$ cases in figure \ref{fig:alphat}. These final state populations are in good agreement with those found by numerically integrating the differential equations for select values of $\alpha$ shown by the black crosses. The most pronounced difference to the Hermitian case is of course visible in the adiabatic limit. 

 While the overall structure of (\ref{eqn:P_0knHerm}) and (\ref{eqn:M_0m}) are similar, the pronounced difference stems form the difference already discussed in the $2\times2$ case, where in the Hermitian system in the adiabatic limit $A\to1$ and $B\to0$, where as in the non-Hermitian system $A$ and $B$ tend to infinity at the same rate. This results in the splitting of the population into equal parts when traversing through the exceptional point in the $2\times2$ system. In combination with the binomial coefficients arising from the $su(2)$ structure this leads to a nontrivial population ratio when the system is driven adiabatically through the exceptional points of higher order.  

In the adiabatic limit the transition matrix elements become
\begin{equation}
\begin{split}
M_{jk}^{(ad)}\to& \frac{A^{n}}{2^{n}}\sum_{l,l'}\begin{pmatrix}k\\l \end{pmatrix}\begin{pmatrix}n-k\\j-l \end{pmatrix}\begin{pmatrix}n-j\\k-l' \end{pmatrix}\begin{pmatrix}j\\l' \end{pmatrix}\\
&=\frac{A^{n}}{2^{n}}\sum_{l}\begin{pmatrix}k\\l \end{pmatrix}\begin{pmatrix}n-k\\j-l \end{pmatrix}\sum_{l}\begin{pmatrix}j\\l \end{pmatrix}\begin{pmatrix}n-j\\k-l \end{pmatrix}.
\end{split}
\end{equation}
Using the Chu-Vandermonde identity
\begin{equation}
\sum_{l}\begin{pmatrix}k\\l \end{pmatrix}\begin{pmatrix}n-k\\j-l \end{pmatrix}=\begin{pmatrix}n\\j \end{pmatrix},
\end{equation}
this reduces to
\begin{equation}
M_{jk}^{(ad)}\to \frac{A^{n}}{2^{n}}\begin{pmatrix}n\\j \end{pmatrix}\begin{pmatrix}n\\k \end{pmatrix}.
\end{equation}
Thus, we have 
\begin{equation}
\sum_l M_{jl}^{(ad)}\to \frac{A^{n}}{2^n} \begin{pmatrix}n\\j \end{pmatrix}\sum_l \begin{pmatrix}n\\l  \end{pmatrix} =A^{n}\begin{pmatrix}n\\j \end{pmatrix}.
\end{equation}
That is, in the adiabatic limit the transition probabilities between the asymptotic eigenstates are given by the weights of the respective binomial coefficients
\begin{equation}
P_{jk}\to \frac{1}{2^n} \begin{pmatrix}n\\k \end{pmatrix},
\end{equation}
rather than a simple equidistribution between the states. 
For the $3\times 3$ case the adiabatic transition probabilities are given by
\begin{equation}
P_{j0} = 
\begin{pmatrix}
\frac{1}{4} & 
\frac{1}{2} &
\frac{1}{4} 
\end{pmatrix}^{T},    
\end{equation}
and in the $4\times 4$ case we have 
\begin{equation}
P_{j0} = 
\begin{pmatrix}
\frac{1}{8} & 
\frac{3}{8} &
\frac{3}{8} &
\frac{1}{8}
\end{pmatrix}^{T},   
\end{equation}
as is clearly visible in figure \ref{fig:alphat}.
Note that this probability does not, in fact, depend on which of the asymptotic eigenstates we start in.

While in the Hermitian case, the adiabatic limit is a trivial consequence of the adiabatic theorem, in the presence of the exceptional points we observe what at first might appear to be a non-intuitive behaviour. In the following we will show that this behaviour, can be in fact understood analytically on the grounds of the adiabatic theorem for non-Hermitian systems, even though non-Hermiticity is often associated with a \textit{breakdown of adiabaticity}.

\section{Adiabatic theorem and adiabatic limit in the non-Hermitian case}
\label{sec:AdiabaticTheorem}
The famous adiabatic theorem of Hermitian quantum mechanics, states that if a system is initially in an eigenstate and the parameters are varied sufficiently slowly, the state will remain arbitrarily close to the instantaneous eigenstates that is smoothly connected to the initial state. The generalisation to the non-Hermitian case has been discussed much in the literature, and leads to some surprising behaviours \cite{Berry2011,Graefe2013,Uzdin2011,Lefebvre2009,Mailybaev2005}. Here we briefly review the derivation of a non-Hermitian version of the adiabatic theorem, which will allow us to correctly deduce the nontrivial adiabatic limit obtained in the LZSM problem with Hamiltonian (\ref{eqn:genNHt}).

For this purpose, we will need the concept of left eigenvectors. The left eigenvectors $|\chi_j\rangle$ of an operator $\hat H$ are defined as the right eigenvectors of the adjoint operator $\hat H^\dagger$ according to 
\begin{equation}
    \hat H^\dagger|\chi_j\rangle=E_j^*|\chi_j\rangle,
\end{equation}
that is 
\begin{equation}
    \langle \chi_j|\hat H=\langle\chi_j|E_j.
\end{equation}
While the eigenvectors of non-Hermitian operators are not orthogonal to each other in general, the right eigenvectors are orthogonal to the left eigenvectors belonging to different eigenvalues, i.e., we have the biorthogonality relation $\langle \chi_j|\phi_k\rangle=\alpha_k\delta_{jk}$. There are several normalisation conventions for non-Hermitian systems \cite{Brody2013}, and while a different choice does not influence the results as long as the convention is followed consistently, different choices are more or less convenient for different situations. 
It is often convenient to renormalise the states such that $\alpha_k=1$, using
\begin{equation}
\langle\chi_j|\phi_k\rangle= \delta_{jk}.
\end{equation} 
In addition, however, there remains a further degree of freedom, regarding the Euclidean norm of the left and right eigenvectors, compatible with the biorthonormal convention. If we have a set of left and right eigenvectors $|\tilde \chi_j\rangle$ and $|\tilde\phi_j\rangle$ that fulfil $\langle \tilde\chi_j|\tilde\phi_k\rangle=\delta_{jk}$, then so do the scaled eigenvectors $|\phi_j\rangle=f_j|\tilde\phi_j\rangle$, and $|\chi_j\rangle=\frac{1}{f_j^*}|\tilde\chi_j\rangle$, which leads to different Euclidean norms, $\langle \chi_j|\chi_j\rangle=\frac{1}{|f_j|^2}\langle\tilde\chi_j|\tilde\chi_j\rangle$, and $\langle \phi_j|\phi_j\rangle=|f_j|^2\langle\tilde\phi_j|\tilde\phi_j\rangle$. For our purposes it will be convenient to adopt the convention
\begin{equation}
\langle\phi_j|\phi_j\rangle=\langle\chi_j|\chi_j\rangle.
\label{eqn:biortho_norm}
\end{equation}
This is achieved by the rescaling
\begin{equation}
|\chi_j\rangle\to\left(\frac{\langle\phi_j|\phi_j\rangle}{\langle\chi_j|\chi_j\rangle}\right)^{\frac{1}{4}}\!|\chi_j\rangle,
\end{equation}
and
\begin{equation}
|\phi_j\rangle\to\left(\frac{\langle\chi_j|\chi_j\rangle}{\langle\phi_j|\phi_j\rangle}\right)^{\frac{1}{4}}\!|\phi_j\rangle,
\end{equation}
which fulfil the required normalisation conditions.  For symmetric Hamiltonians, $\hat H^T=\hat H$, the left eigenvectors are proportional to the complex conjugate of the right eigenvectors, $|\chi_j\rangle\propto|\phi_j^*\rangle$ \cite{Keck2003}, which can be simply seen by taking the complex conjugate of the left eigenvalue equation.

To derive an adiabatic theorem, let us begin by considering a set of instantaneous right eigenvectors $|\phi_{j}(t)\rangle$, and left eigenvectors $|\chi_{j}(t)\rangle$ of $\hat H$, i.e.,
\begin{equation}
\hat{H}|\phi_{j}(t)\rangle = E_{j}(t)|\phi_{j}(t)\rangle,
\label{eqn:eigenval}
\end{equation}
and 
\begin{equation}
\hat{H}^\dagger|\chi_{j}(t)\rangle = E_{j}^{*}(t)|\chi_{j}(t)\rangle.
\end{equation}
For convenience we assume these to be normalised according to the biorthogonal inner product $\langle\chi_{j}(t)|\phi_{k}(t)\rangle = \delta_{jk}$. 
Away from the exceptional points the $|\phi_{j}(t)\rangle$ form a complete basis, and we can express the time-dependent state as  
\begin{equation}
|\psi(t)\rangle = \sum_{j}c_{j}(t)|\phi_{j}(t)\rangle,
\label{eqn:expansion}
\end{equation}
with $c_j(t)=\langle \chi_j(t)|\psi(t)\rangle$. For ease of notation we omit the explicit time dependence in the following calculation. Inserting (\ref{eqn:expansion}) into the Schr\"{o}dinger equation yields
\begin{equation}
i\sum_{j}\left(\dot c_{j}|\phi_{j}\rangle+c_{j}|\dot \phi_{j}\rangle\right)=\sum_{j}c_{j}E_{j}|\phi_{j}\rangle, 
\end{equation}
which when projected onto the left eigenstates yields
\begin{equation}
i\dot{c}_{j} = (E_{j}-i\langle \chi_{j}|\dot\phi_{j}\rangle)c_{j} - i\sum_{k\neq j}\langle \chi_{j}|\dot{\phi}_{k}\rangle c_{k}.
\label{eqn:ad_cdot1}
\end{equation}
Since our Hamiltonian is symmetric, i.e. $\hat H=\hat H^T$, we can choose 
\begin{equation}
\langle \chi_j|=\langle\phi_j^*|,
\end{equation} 
and differentiating the biorthogonal normalisation condition $\langle\chi_j|\phi_j\rangle=1$ then yields
\begin{equation}
\langle \dot\phi_j^*|\phi_j\rangle+\langle \phi_j^*|\dot\phi_j\rangle=2\langle \phi_j^*|\dot\phi_j\rangle=2\langle \chi_j|\dot\phi_j\rangle=0,
\end{equation}
Thus, the diagonal term in (\ref{eqn:ad_cdot1}) simplifies and we have 
\begin{equation}
i\dot{c}_{j} = E_{j}c_{j} - i\sum_{k\neq j}\langle \chi_{j}|\dot{\phi}_{k}\rangle c_{k}.
\label{eqn:cdot}
\end{equation}
If the eigenvalues $E_{j}$ are real, we can argue just as in the Hermitian case, that if the parameters are varied sufficiently slowly, as long as the energies remain non-degenerate, the non-diagonal terms vanish. This can be seen explicitly by taking the time derivative of the eigenvalue equation (\ref{eqn:eigenval}) and projecting it onto the left eigenstate to find 
\begin{equation}
\langle \chi_{j}|\dot{\phi}_{k}\rangle = \frac{\langle \chi_{j}|\dot{\hat{H}}|\phi_{k}\rangle}{E_{k}-E_{j}},
\label{Adiabatic theorem}
\end{equation}
which indeed goes to zero for sufficiently slow parameter variation, in our case for $\alpha \rightarrow 0$. Thus, in the adiabatic limit, away from degeneracies, equation (\ref{eqn:cdot}) reduces to
\begin{equation}
\dot{c}_{j} = -i E_{j}c_{j}.
\end{equation}
If the $E_{j}$ are complex with different imaginary parts, on the other hand, not only do the relative populations change due to the exponential time dependence with different rates, but also the argument to neglect the non-adiabatic coupling elements in (\ref{eqn:cdot}) has to be treated more carefully, as even exponentially small transitions can be amplified to become dominant contributions by the relative exponential growth between different instantaneous states. This effect is sometimes referred to as the ``breakdown of the adiabatic theorem'' and indeed it can lead to a derivation from the naive expectation that a system follows an instantaneous eigenstate if driven adiabatically \cite{Doppler2016}.

Of course we cannot expect to be  able to apply the adiabatic theorem directly when considering a system driven through an exceptional point, since the eigenvectors are not analytic functions of the parameters at the exceptional point and thus, it is meaningless to try to connect eigenfunctions on one side of the exceptional point to those on the other side. In our case, however, we can nevertheless apply the adiabatic theorem to understand the splitting of the population we have observed in the LZSM result, as we shall explain in what follows. 

Due to the different imaginary parts of the eigenvalues between the two exceptional points, the time-dependent state in the adiabatic case is given by the instantaneous eigenstate with the largest imaginary part between the two exceptional points. As the second exceptional point is approached, this state becomes the single exceptional point eigenstate. Thus, we can assume that at the time at which the exceptional point $t=t_{EP}^{+}$ is reached, the system is in the state $|\phi_{EP}\rangle=|n_y\rangle$. After the EP the eigenstates change in a non-analytic fashion, with $N$ distinct eigenstates emerging from the single exceptional point eigenstate. For an infinitesimally short time after the EP, however, the state itself cannot have changed, as the time-evolution governed by the Schr\"odinger equation is smooth. Thus, we can assume that 
for an infinitesimally small time after $t_{EP}^{+}$, $t_{EP}^{+}+\delta t$, the wavefunction is still given by $|\psi(t_{EP}^{+}+\delta t)\rangle=|\phi_{EP}\rangle$. At this point, there exists a complete (highly non-orthogonal) basis of eigenstates again, in which the state can be expanded as in equation (\ref{eqn:expansion}), with 
\begin{equation}
\begin{split}
|c_{j}(t_{EP}^{+} +\delta t)|^2 &= |\langle \chi_{j}(t_{EP}^{+}+\delta t)|\psi(t_{EP}^{+}+\delta t)\rangle|^2\\
&=|\langle \chi_{j}(t_{EP}^{+}+\delta t)|\phi_{EP}\rangle|^2.
\end{split}
\end{equation}

If the parameter variation continues in an adiabatic fashion, the $|c_{j}|^2$ will remain constant for the remainder of the time evolution, and we have 
\begin{equation}
|\psi(t)\rangle = \sum_{j}c_{j}(t_{EP}^{+}+\delta t)\rme^{-\rmi\int_{t_{EP}^+}^t E(t')\rmd t'}|\phi_{j}(t)\rangle.
\end{equation}
In the long time limit the $|\phi_j\rangle$ are given by the orthonormal $\hat J_z$ eigenvectors, and we can interpret the renormalised $|c_j|^2$ as transition probabilities 
\begin{equation}
P_{j}(t\!\to\!\infty)\!=\frac{|c_{j}(t_{EP}^{+}+\delta t)|^2}
{\sum_{l}|c_{l}(t_{EP}^{+}+\delta t)|^2},
\end{equation}
for infinitesimally small $\delta t$. 

To calculate the left eigenvectors of the Hamiltonian (\ref{eqn:genNHt}), instead of using the relation between left and right eigenvectors explicitly, we write
\begin{equation}
\hat{H}^\dagger= 2\lambda^*\,\hat X\hat{J}_{y}\hat X^{-1},   
\label{eqn-Hdag_diagy}
\end{equation}
with 
\begin{equation}
\hat X=\rme^{\rmi\frac{\pi}{2}\frac{1}{\lambda}(\alpha t\hat J_x-\rmi\gamma\hat J_z)}.
\end{equation}
Of course it holds $\hat X^{-1}=\hat \Phi^\dagger$. Thus, the left eigenvectors are up to a factor given by
\begin{equation}
|\chi_j\rangle\propto\hat X|j_y\rangle.
\label{eqn:chi_m}
\end{equation}

Working with the normalisation convention (\ref{eqn:biortho_norm}) in what follows we use the rescaled eigenvectors
\begin{equation}
|\phi_j\rangle=\left(\frac{\langle j_y|\hat X^\dagger\hat X|j_y\rangle}{\langle j_y|\hat\Phi^\dagger\hat \Phi|j_y\rangle}\right)^{\frac{1}{4}}\hat\Phi|j_y\rangle,
\label{eqn:n_phi_m}
\end{equation}
and 
\begin{equation}
|\chi_j\rangle=\left(\frac{\langle j_y|\hat \Phi^\dagger\hat \Phi|j_y\rangle}{\langle j_y|\hat X^\dagger\hat X|j_y\rangle}\right)^{\frac{1}{4}}\hat X|j_y\rangle.
\label{eqn:n_chi_m}
\end{equation}

Using this we have
\begin{equation}
|c_{j}|^2 = |\langle j_y|\hat X^\dagger|n_y\rangle|^2\left(\frac{\langle j_y|\hat \Phi^\dagger\hat \Phi|j_y\rangle}{\langle j_y|\hat X^\dagger\hat X|j_y\rangle}\right)^{\frac{1}{2}}.
\label{eqn:c_j_overlap}
\end{equation}

To deduce the relevant matrix elements in the $N\times N$ representation we again use equation (\ref{eqn:D}). We start from the 
 matrix representations of $\hat X$ and $\hat \Phi$ in the $2\times 2$ case (in the $\hat J
 _y$ basis), which are given by
\begin{equation}
\hat X^{(2)}= \frac{1}{\sqrt{2}}\begin{pmatrix}  1 & -x^{-1}\\ x & 1\end{pmatrix}, 
\end{equation}
and
\begin{equation}
\hat\Phi^{(2)}= \frac{1}{\sqrt{2}}\begin{pmatrix}  1 & -x\\ x^{-1}& 1\end{pmatrix}, 
\end{equation}
where we have introduced the abbreviation
\begin{equation}
x=\frac{\gamma-\alpha t}{\lambda}.
\end{equation}
That is, we have 
\begin{equation}
(\hat\Phi^\dagger\hat \Phi)^{(2)}= \frac{1}{2}\begin{pmatrix}  1+x^{-2} & -x+x^{-1}\\ -x+x^{-1}& 1+x^2\end{pmatrix}, 
\end{equation}
and
\begin{equation}
(\hat X^\dagger\hat X)^{(2)}= \frac{1}{2}\begin{pmatrix}  1+x^2 & x-x^{-1}\\ x-x^{-1}& 1+x^{-2}\end{pmatrix}.
\end{equation}

From this we find

\begin{equation}
\langle j_y|\hat \Phi^\dagger\hat \Phi|j_y\rangle=x^{2(2j-n)}\langle j_y|\hat X^\dagger\hat X|j_y\rangle,
\end{equation}
and thus equation (\ref{eqn:c_j_overlap}) simplifies to
\begin{equation}
\nonumber |c_{j}|^2 = x^{2j-n} |\hat X^\dagger_{j,n}|^2.
\end{equation}

Using equation (\ref{eqn:D}) for $\hat X^\dagger$ we further find 
\begin{equation}
\hat X^{\dagger\,(N)}_{jn}=\frac{x^{n\!-\!j}}{\sqrt{2}^n}\begin{pmatrix}n\!\\j\!\end{pmatrix}^{\!\!\!-\frac{1}{2}},
\end{equation}
which reduces (\ref{eqn:c_j_overlap}) to the simple expression
\begin{equation}
|c_{j}|^2 =   \begin{pmatrix}n\!\\j\!\end{pmatrix}\frac{x^n}{2^n}, 
\end{equation}
and thus 
\begin{equation}
\sum_k |c_{k}|^2 =  \frac{x^n}{2^n}\sum_k \begin{pmatrix}n\!\\k\!\end{pmatrix}=x^n. 
\end{equation}
Thus, we finally find for the asymptotic transition probability 
\begin{equation}
P_{j}(t\to\infty)=\frac{1}{2^n}\begin{pmatrix}n\!\\j\!\end{pmatrix},
\end{equation}
in agreement with the result from the LZSM calculation. 

That is, the splitting of population at the exceptional point, can indeed be understood as an adiabatic effect where the ratios of population in the different branches are given by the overlaps of the relevant eigenstates with the exceptional point state. 

\section{Summary and Outlook}
\label{sec:sum}
In summary we have derived the full set of transition probabilities between the asymptotic eigenstates for an $N$-level, non-Hermitian, PT-symmetric, Landau-Zener-St\"uckelberg-Majorana problem with two exceptional points of order $N$. In the adiabatic limit the final populations are given by binomial coefficients. We have provided an analytical argument based on the adiabatic theorem and the structure of the instantaneous eigenvectors to derive the result in the adiabatic limit independently. 

It is an interesting question how the transfer probabilities would change in a system with $N$-th order exceptional points with a different unfolding pattern. The $SU(2)$ case discussed here, lends itself naturally to the analysis of the transition probabilities, as the energies are completely real in both asymptotic regimes $t\to\pm\infty$. For an exceptional point of order $N$ to unfold into a purely real set  of eigenvalues in one direction of parameter variation in  fact implies either a square root or a linear unfolding. Thus, for other unfolding patterns the problem could not be considered in the asymptotic limit and a different approach would be needed. Another interesting question concerns transitions in a series of exceptional points of lower order, in which interference effects between different transition branches would play a role. 

\section*{Acknowledgements}
E.M.G. and S.M. acknowledge  support  from from the European Research Council (ERC) under the European Union's Horizon 2020 research and innovation program (grant agreement No 758453), E.M.G acknowledges support from the Royal Society (Grant. No. URF\textbackslash R\textbackslash 201034), and R.M. acknowledges support from an IC President's PhD scholarship . 

\appendix
\section{Deducing $N\times N$ representations of $SL(2)$ group elements from the $2\times 2$ representation}
\label{app:SL2}

In this appendix, for completeness we provide a brief derivation of equation (\ref{eqn:D}) which is central to many of the calculations in this paper. The main idea of the derivation uses the basis of $SU(2)$ coherent states (which are in fact, equivalent to the set of $SL(2)$ coherent states), often referred to as the spinor representation in the literature, to deduce the $N$ dimensional basis representation of a group element \cite{Hammermesh1963}. 

An $SU(2)$ coherent state in $N$ dimensions can be represented as a vector with components 
\begin{equation}
 \psi_j^{(N)} = \sqrt{\begin{pmatrix}n \\ j \end{pmatrix}}(\psi_{1})^{n-j}(\psi_{2})^{j},
\label{eqn:Nlevel}
\end{equation}
in the standard basis, with $\psi_{1,2}\in\mathds{C}$, where $n=N-1$, and $j$ runs from $0$ to $n$. The spin $\frac{1}{2}$ representation is simply given by $\psi^{(2)}=(\psi_{1},\psi_{2})^{T}$. 

To deduce the $N$ dimensional representation $\hat D^{(N)}$ from the $2$ dimensional one, which we denote by $\hat d$,  we use the fact that coherent states are mapped into coherent states by an $SL(2)$ operator, and solve the linear system of equations 
\begin{equation}
\hat D^{(N)} \psi^{(N)}=\tilde \psi^{(N)},
\end{equation}
for $\hat D$, 
with
\begin{equation}
\tilde\psi^{(2)}=\hat d\, \psi^{(2)}=\begin{pmatrix}\tilde\psi_{1}\\ \tilde\psi_{2}\end{pmatrix}=\begin{pmatrix} d_{11}\psi_{1}+d_{12}\psi_{2} \\ d_{21}\psi_{1} + d_{22}\psi_{2} \end{pmatrix}. 
\label{eqn:map2}
\end{equation}
Explicitly we have 
\begin{equation}
\begin{split}
\tilde\psi_j^{(N)} &= \sqrt{\begin{pmatrix}n \\ j \end{pmatrix}}(d_{11}\psi_{1}+d_{12}\psi_{2} )^{n-j}(d_{21}\psi_{1} + d_{22}\psi_{2})^{j}\\
&=\sum_{k=0}^{n-j}\sum_{l=0}^{j}\alpha_{k,l}^{(N,j)} \psi_{1}^{n-k-l} \psi_{2}^{k+l},
\label{eqn:longRHS}
\end{split}
\end{equation}
with
\begin{equation}
\alpha_{k,l}^{(N,j)}= \begin{pmatrix}n \\ j \end{pmatrix}^{\frac{1}{2}}\begin{pmatrix}j \\ l \end{pmatrix}\begin{pmatrix}n-j \\ k \end{pmatrix}d_{11}^{n-j-k} d_{12}^{k} d_{21}^{j-l} d_{22}^{l}.
\end{equation}
Whereas by definition we have 
\begin{equation}
\begin{split}
\left(\hat D^{(N)}\psi^{(N)}\right)_j&=\sum_{m=0}^n (\hat D^{(N)})_{jm} \psi^{(N)}_m\\
&=\sum_{m=0}^{n} \sqrt{\begin{pmatrix}n \\ m \end{pmatrix}} (\hat D^{(N)})_{jm}\psi_{1}^{n-m}\psi_{2}^{m}. 
\label{eqn:shortRHS}
\end{split}
\end{equation}
Rephrasing the sum over $k$ in equation (\ref{eqn:longRHS}) as a sum over $m=k+l$, which runs from $0$ to $n$, the sum over $l$ runs from $l_{min}=\text{max}(m-(n-j),0)$ to $l_{max}=\text{min}(m,j)$, and we have
\begin{equation}
\tilde\psi_j^{(N)}\!=\sqrt{\!\begin{pmatrix}n \\ j \end{pmatrix}\!}\sum_{m=0}^{n}\!\sum_{l_{min}}^{l_{max}}\!\!\begin{pmatrix}j \\ l \end{pmatrix}\!\!\begin{pmatrix}\!n\!-\!j\! \\\! m\!-\!l\! \end{pmatrix}\!d_{11}^{n\!-\!j\!-\!m\!+\!l} d_{12}^{m\!-\!l} d_{21}^{j\!-\!l} d_{22}^{l} \psi_{1}^{n\!-\!m} \psi_{2}^{m}.
\label{eqn:longRHS2}
\end{equation}
Comparing the coefficients in equations  (\ref{eqn:shortRHS}) and  (\ref{eqn:longRHS2}) then yields the expression for the matrix elements of $\hat D$ in the $N$ dimensional representation as
\begin{align}
\nonumber
D^{(N)}_{j,m}\!=&\!\begin{pmatrix}n\!\\j\!\end{pmatrix}^{\!\!\frac{1}{2}}\!\!\!\begin{pmatrix}n\!\\m\!\end{pmatrix}^{\!\!\!-\frac{1}{2}}\!\!\!\sum_{l=l_{\rm min}}^{l_{\rm max}}\!\!\!\begin{pmatrix}n\!-\!j\\m\!-\!l\end{pmatrix}\!\!\begin{pmatrix}j\\l\end{pmatrix}\!d_{11}^{n\!-\!j\!-\!m\!+\!l}d_{12}^{m\!-\!l}d_{21}^{j\!-\!l}d_{22}^{l},
\\
=&\!\!\!\sum_{l=l_{\rm min}}^{l_{\rm max}}\!\!\!\!\tfrac{\sqrt{m!(n\!-\!m)!(n\!-\!j)!j!}}{(n\!-\!j\!-\!m\!+\!l)!(m\!-\!l)!(j\!-\!l)!l!}d_{11}^{n\!-\!j\!-\!m\!+\!l}d_{12}^{m\!-\!l}d_{21}^{j\!-\!l}\!d_{22}^{l}.
\label{eqn:D-app}
\end{align}

\end{document}